\documentclass[aps,showpacs,twocolumn,superscriptaddress,prb]{revtex4-1}
\usepackage{hyperref}
\usepackage[pdftex]{graphicx}
\usepackage{natbib}
\usepackage{epstopdf}
\usepackage{graphicx}
\usepackage{amsmath,amssymb,amsfonts}

\begin{document}

\title{Magnetic and transport properties of i-$R$-Cd icosahedral quasicrystals ($R$ = Y, Gd-Tm)}
\author{Tai Kong}  
\affiliation{Ames Laboratory, U.S. DOE, Iowa State University, Ames, Iowa 50011, USA\\
Department of Physics and Astronomy, Iowa State University, Ames, Iowa 50011, USA}
\author{Sergey L. Bud'ko}  
\affiliation{Ames Laboratory, U.S. DOE, Iowa State University, Ames, Iowa 50011, USA\\
Department of Physics and Astronomy, Iowa State University, Ames, Iowa 50011, USA}
\author{Anton Jesche}  
\affiliation{Ames Laboratory, U.S. DOE, Iowa State University, Ames, Iowa 50011, USA\\
Department of Physics and Astronomy, Iowa State University, Ames, Iowa 50011, USA}
\author{John McArthur}
\affiliation{Quantum Design Japan, 1-11-16 Takamatsu, Toshima ku, Tokyo, 171-0042, Japan}
\author{Andreas Kreyssig}  
\affiliation{Ames Laboratory, U.S. DOE, Iowa State University, Ames, Iowa 50011, USA\\
Department of Physics and Astronomy, Iowa State University, Ames, Iowa 50011, USA}
\author{Alan I. Goldman}  
\affiliation{Ames Laboratory, U.S. DOE, Iowa State University, Ames, Iowa 50011, USA\\
Department of Physics and Astronomy, Iowa State University, Ames, Iowa 50011, USA}
\author{Paul C. Canfield}
\affiliation{Ames Laboratory, U.S. DOE, Iowa State University, Ames, Iowa 50011, USA\\
Department of Physics and Astronomy, Iowa State University, Ames, Iowa 50011, USA}

\begin{abstract}
We present a detailed characterization of the recently discovered i-$R$-Cd ($R$ = Y, Gd-Tm) binary quasicrystals by means of x-ray diffraction, temperature-dependent dc and ac magnetization, temperature-dependent resistance and temperature-dependent specific heat measurements. Structurally, the broadening of x-ray diffraction peaks found for i-$R$-Cd is dominated by frozen-in phason strain, which is essentially independent of $R$. i-Y-Cd is weakly diamagnetic and manifests a temperature-independent susceptibility. i-Gd-Cd can be characterized as a spin-glass below 4.6 K via dc magnetization cusp, a third order non-linear magnetic susceptibility peak, a frequency-dependent freezing temperature and a broad maximum in the specific heat. i-$R$-Cd ($R$ = Ho-Tm) is similar to i-Gd-Cd in terms of features observed in thermodynamic measurements. i-Tb-Cd and i-Dy-Cd do not show a clear cusp in their zero-field-cooled dc magnetization data, but instead show a more rounded, broad local maximum. The resistivity for i-$R$-Cd is of order 300 $\mu \Omega$ cm and weakly temperature-dependent. The characteristic freezing temperatures for i-$R$-Cd ($R$ = Gd-Tm) deviate from the de Gennes scaling, in a manner consistent with crystal electric field splitting induced local moment anisotropy. 
\end{abstract}
\maketitle

\DeclareGraphicsExtensions{.jpg,.pdf,.mps,.png,.eps}
\section{Introduction}
Since the initial discovery of quasicrystals\cite{Shechtman1984}, the search for new quasicrystalline systems, especially thermodynamically stable ones, as well as understanding of their structural and physical properties has been of keen interest to the solid state physics and chemistry communities\cite{Janot1997}. At the expense of losing translational symmetry, rotational symmetries that were forbidden by conventional crystallography, like 5-fold rotational symmetry, could be achieved in quasicrystals. Interestingly, despite several theoretical predictions\cite{Lifshitz98, Wessel03,Vedmedenko04}, no long range magnetic ordering has yet been discovered in moment-bearing quasicrystals. Until recently, this lack of long-range magnetic ordering also extended to quasicrystal approximants\cite{Swenson02,Heggen10,Soshi11}, which can be viewed as quasicrystalline clusters sitting on a periodic lattice that possesses a translational symmetry. Recently, two exceptions have been identified: ferromagnetic Gd-Au-Si(Ge)\cite{Hiroto13} and antiferromagnetic $R$Cd$_{6}$\cite{Tamura10,Kazue13,Mori12,Kim12,Kreyssig13}. The antiferromangetic $R$Cd$_{6}$ compounds in particular have attracted great attention, since they bring up the possibility of related quasicrystal phases that could have long-range magnetic ordering. Additionally, the $R$Cd$_{6}$ series offers an opportunity to look into how magnetism evolves from a conventional lattice (quasicrystal approximant phase) to an aperiodic quasicrystal. However, the previously discovered corresponding binary quasicrystals, YbCd$_{5.7}$\cite{tsai} and CaCd$_{5.7}$\cite{Guo00}, do not bear local moments.

Recently, based upon the idea that quasicrystals may exist near approximant phases as compounds with relatively low peritectic decomposition temperatures\cite{Canfield10}, a new, stable binary quasicrystal phase, i-$R$-Cd for $R$ = Y, Gd-Tm was discovered\cite{QC2013}. In order to better understand the i-$R$-Cd quasicrystalline series, in this paper we detail structural characterization by x-ray diffraction as well as thermodynamic and transport characterization of i-$R$-Cd ($R$ = Y, Gd-Tm). 
  
\section{Experimental Methods}
\begin{figure}[tbh!]
\includegraphics[scale = 0.35]{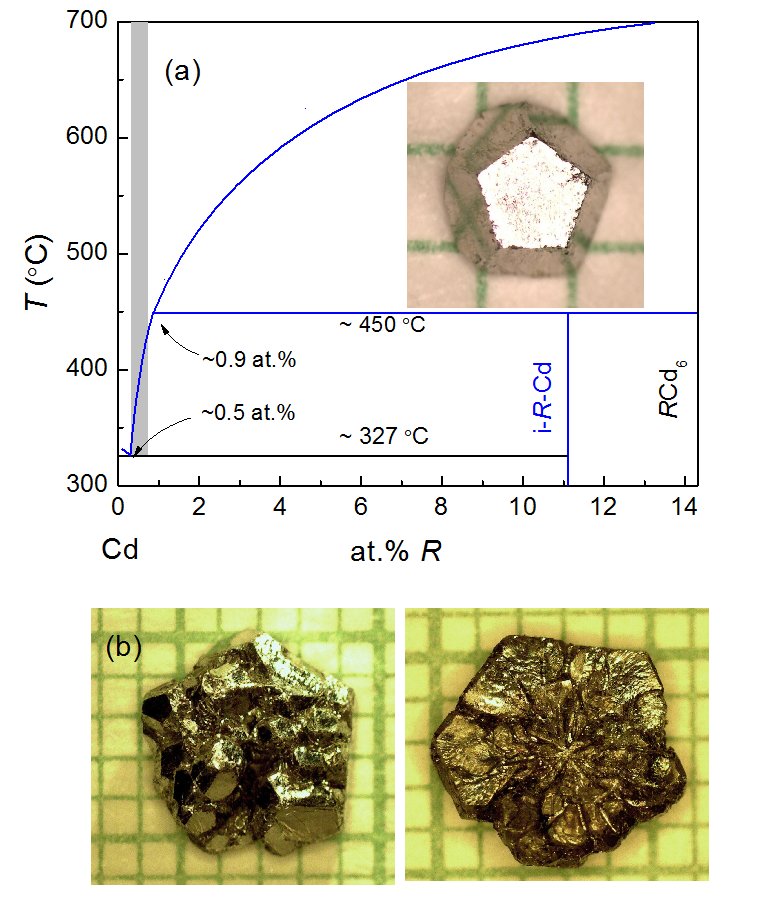}
\caption{(Color online) (a) Generalized $R$-Cd binary phase diagram around the Cd concentrated region. The inset shows a single grain of i-Gd-Cd on a millimeter grid paper. (b) The front and back sides of a larger grain of i-Tb-Cd on a millimeter grid paper.}
\label{diagram}
\end{figure}

Quasicrystals were grown from a binary melt using a solution growth method\cite{QC2013,Canfield01}. Fig.~\ref{diagram}(a) presents a generic $R$-Cd ($R$ = Y, Gd-Tm) binary phase diagram, in which the grey region indicates the composition that allows primary solidification of the quasicrystalline phase. Typical starting compositions are $R$:Cd = 0.8:99.2 for $R$ = Y, Gd-Dy and 0.6:99.4 for $R$ = Ho-Tm. The starting elements were put into an alumina crucible and sealed in a quartz ampoule that was then heated up to 700 $^{\circ}$C and slowly cooled to 330 $^{\circ}$C, at which temperature the remaining solution was decanted. Despite several attempts, i-$R$-Cd for $R$ = Nd, Sm, Yb and Lu could not be grown. In Fig.~\ref{diagram}, the typical habits of i-$R$-Cd quasicrystal are shown.  Small, single grains of quasicrystal, like the one shown in the inset of Fig.~\ref{diagram}(a), usually are well faceted. Much larger grains, as shown for both the front and back sides in Fig.~\ref{diagram}(b), often form from a single nucleation site, which then follows an initial dendritic growth and followed by faceted growth.

Powder x-ray diffraction patterns were measured on a Rigaku Miniflex II desktop x-ray diffractometer using Cu K$\alpha_{1,2}$ radiation at ambient temperature. Samples were prepared by grinding single grains into a powder that was then placed on a Si single-crystal, zero-background, sample holder.  Data were taken using steps of 0.01$^\circ$ in the scattering angle, 2$\theta$, counting for 6 seconds at each step. High-energy x-ray diffraction data were collected at the Advanced Photon Source, Argonne National Laboratory.

dc magnetization data down to 2 K were measured using a Quantum Design (QD) Magnetic Property Measurement System (MPMS), Superconducting Quantum Interference Device (SQUID) magnetometer ($T$ = 1.8-300 K, $H_\text{max}$ = 55 kOe). dc magnetization below 2 K was measured at QD (Japan) using an iHelium3 system.

ac magnetic susceptibility, specific heat and resistance were measured using a QD Physical Property Measurement System (PPMS). To obtain the non-linear magnetic susceptibility, $\chi_{3}$, the real part of magnetic susceptibility was measured with a biased dc field ranging from -500 Oe to 500 Oe and was fit with non-linear terms at each temperature. Specific heat was measured using a QD PPMS via relaxation method. A $^{3}$He option was utilized to enable the measurements down to 0.4 K. Without an extrapolation of specific heat down to 0 K, the estimate of magnetic entropy starts from 0.4 K. Resistance was measured using a standard 4-probe, ac technique ($f$ = 17 Hz, $I$ = 3 mA). Epotek-H20E silver epoxy was used to attach Pt wires onto the sample. Although resistance samples were polished into rectangular bars that allow for resistivity measurement, only normalized resistance will be presented as a result of potential elemental Cd contamination in the sample. This will be discussed in detail in the Appendix. In general, the resistivity of i-$R$-Cd is about 300 $\mu\Omega$ cm at room temperature and only weakly temperature-dependent.

\section{Results and Analysis}

\subsection{x-ray diffraction}

\begin{figure}
\centering
\includegraphics[scale = 0.5]{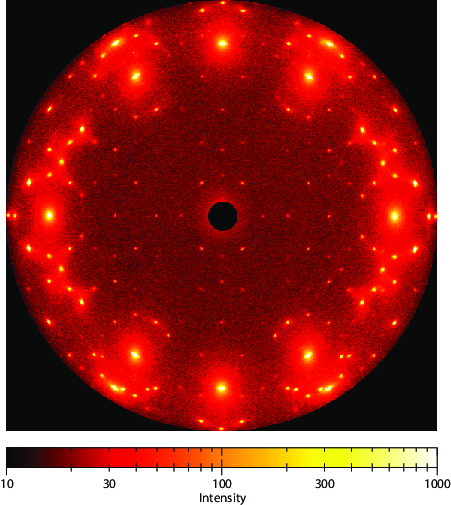}
\caption{(Color online) High-energy x-ray diffraction pattern of the i-Tb-Cd quasicrystal shown in Fig.~\ref{diagram}(b). The recorded reciprocal plane is perpendicular to the 2-fold direction of the icosahedral quasicrystal. The logarithmic intensity scale emphasizes weak signals relative to the strongest Bragg peaks with maximum counts up to 146,000.}
\label{hexray}
\end{figure}

In Fig.~\ref{diagram}(b), the large grain of a single phase quasicrystal does not appear to preserve the ideal single pentagonal dodecahedron faceting. Although there was an initial dendritic growth, the whole grain that results from a single nucleation site does maintain a single orientation. The same i-Tb-Cd quasicrystal as shown in Fig.~\ref{diagram}(b) was studied on the instrument 6-ID-D at the Advanced Photon Source, Argonne National Laboratory using 100.3 keV x-ray giving an absorption length of approximately 0.8 mm and, therefore allowing full penetration of the sample. Two-dimensional scattering patterns were measured by a MAR345 image plate positioned 2814 mm behind the sample. Entire reciprocal planes have been recorded using the method described in detail in Ref.~\onlinecite{Kreyssig07} by tilting the sample perpendicular to the incident x-ray beam by $4.0^{\circ}$ through two independent angles, $\mu$ and $\eta$. A typical diffraction pattern is shown in Fig.~\ref{hexray} with the recorded reciprocal plane perpendicular to the 2-fold direction of the icosahedral quasicrystal. In each exposure, a large sample volume was probed defined by the beam dimensions of 1$\times$1 mm$^{2}$ confined by the incident beam slit system and the full width of the sample along the beam direction. The entire sample has been surveyed by recording a series of diffraction patterns and translating the sample in both directions perpendicular to the beam in a grid-like manner in steps of 1 mm. All recordings showed similar patterns demonstrating the same crystal orientation in each probed sample volume and, therefore, demonstrating that the entire sample shown in Fig.~\ref{diagram}(b) is a single grain i-Tb-Cd quasicrystal despite the obvious initial dendritic growth. Only minor traces of Cd flux have been detected as impurity phases. A measurement of a second i-Tb-Cd sample yielded similar results.

\begin{figure}
\centering\includegraphics[width=0.95\linewidth]{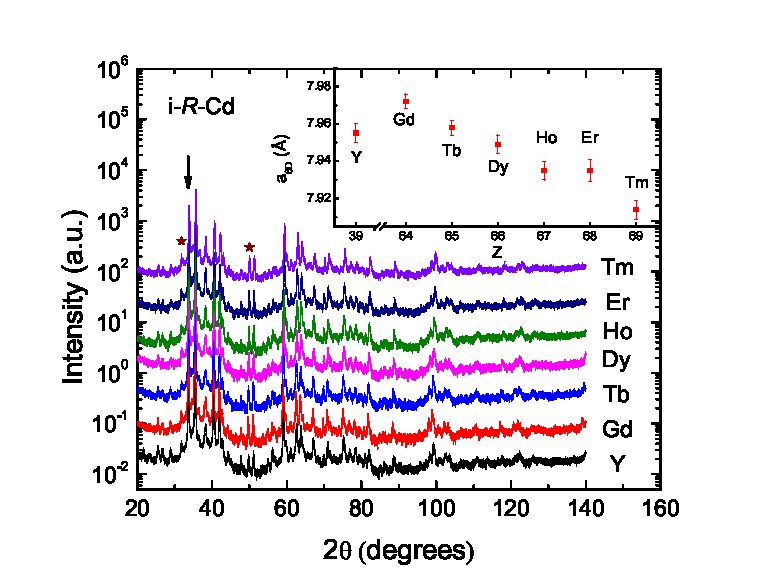}\\
\caption{(Color online) Powder diffraction patterns from all of the i-$R$-Cd quasicrystals under investigation. The patterns are normalized to the strongest diffraction peak and offset for clarity.  All peaks can be indexed to either the icosahedral phase or residual Cd flux. Stars indicate major diffraction peaks that come from Cd flux. The arrow indicates the (211111) peak. The inset shows the 6D quasilattice parameter a$_\text{6D}$ as a function of atomic number, Z, of rare earth, $R$.}
\label{xray}
\end{figure}

The ambient temperature powder x-ray diffraction patterns obtained for i-$R$-Cd ($R$ = Y, Gd-Tm) are shown in Fig.~\ref{xray}. All diffraction peaks from all samples can be indexed to the primitive icosahedral phase pattern and varying small amounts of residual Cd flux. Progressing from $R$ = Gd to $R$ = Tm we see that the peaks shift to slightly higher values of 2$\theta$, reflecting a change in the six-dimensional quasilattice constant, $a_{\rm{6D}}$. Using the strongest peak along the five-fold axis (indexed (211111) in Ref.~\onlinecite{QC2013} and indicated by an arrow in Fig.~\ref{xray}), $a_{\rm{6D}}$ can be calculated to range from 7.972(4) \AA~ for i-Gd-Cd to 7.914(5) \AA~ for i-Tm-Cd. The quasilattice parameter of i-Y-Cd is close to those of i-Tb-Cd and i-Dy-Cd.

As discussed in Ref.~\onlinecite{QC2013}, there may be a slight change in stoichiometry as $R$ changes from Gd to Tm. In order to see to what extent this leads to the changes in structural disorder, we evaluated the degree of phason strain in the two structural extremes, $R$ =  Gd and Tm. Phason disorder and frozen-in phason strain arise in aperiodic systems as a result of additional degrees of freedom in density wave descriptions of quasicrystals or can be viewed in terms of flips or errors in the tiling description of aperiodic systems \cite{Lubensky_1988}. For a general description of phason strain in quasicrystals we refer the reader to Ref.~\onlinecite{Widom08} and references therein. For our purposes here, it is sufficient to note that phason strain translates to disorder in the atomic scale structure.

\begin{figure}[tbt!]
\centering\includegraphics[width=1\linewidth]{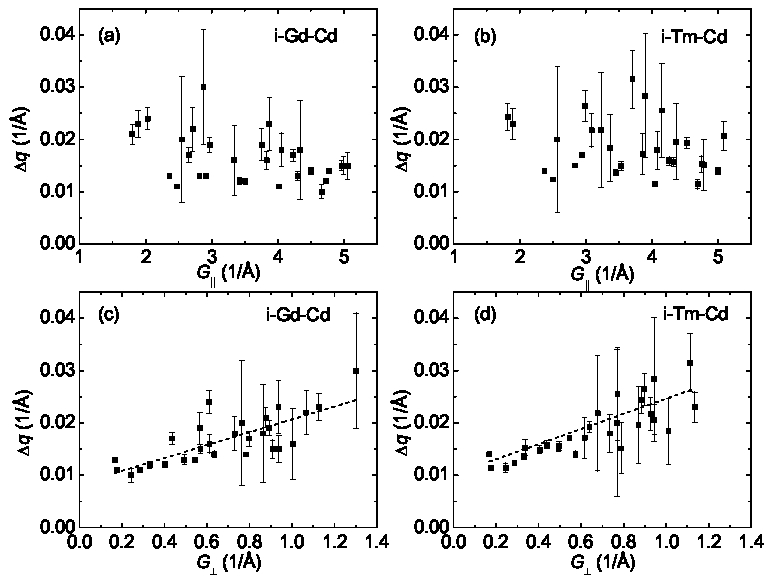}\\
\caption{Systematics of diffraction peak broadening in i-Gd-Cd and i-Tm-Cd which represent the extremes in the range of $R$ concentration. Panels (a) and (b) plot the diffraction peak widths for each compound vs. $G_{\rm{\|}}$, and panels (c) and (d) plot the diffraction peak widths for each compound vs. $G_{\rm{\bot}}$ as described in the text.  The dashed lines represent the best fit straight lines to the data.}\label{phason_strain}
\end{figure}

The presence of frozen-in phason strain within a given quasicrystalline sample is signalled by a systematic broadening of diffraction peaks and/or shifts of diffraction peak positions and/or the presence of diffuse scattering. Unlike physical strain, which results in diffraction peak broadening that scales with the physical momentum transfer, $G_{\rm{\|}}$, phason strain broadening of diffraction peaks scales with the perpendicular space momentum, denoted $G_{\rm{\bot}}$ \cite{Socolar87}. Each peak in the diffraction patterns shown in Fig.~\ref{xray} can be associated with distinct values for $G_{\rm{\|}}$ and $G_{\rm{\bot}}$ and, in Fig.~\ref{phason_strain}, we plot the width of diffraction peaks as a function of their values of $G_{\rm{\|}}$ and $G_{\rm{\bot}}$ for $R$ = Gd [Figs.~\ref{phason_strain}(a) and (c)] and $R$ = Tm [Figs.~\ref{phason_strain}(b) and (d)]. The widths of the diffraction peaks were determined from fits using a pseudo-Voigt function taking into account both the Cu K$\alpha_{1}$ and K$\alpha_{2}$ contributions to the profile. Whereas the peak broadening for both $R$ = Gd and Tm display no particular trend with $G_{\rm{\|}}$, the essentially linear dependence of peak broadening with $G_{\rm{\bot}}$ indicates that the frozen-in phason strain is the predominant mechanism for peak broadening in these samples. The resolution of the powder diffractometer was measured, using a Si powder standard, to be $\Delta Q \approx$~0.01~\AA~full-width-at-half-maximum.  Therefore, the peaks at the smallest values of $G_{\rm{\bot}}$ are resolution limited.  We also note that the magnitude of the phason strain and its dependence on $G_{\rm{\bot}}$ for i-Gd-Cd and i-Tm-Cd can not be readily distinguished, indicating that the degree of phason strain in these samples is comparable and shows no clear dependence on $R$ (either size or precise concentration).  We further note that the systematics and magnitude of diffraction peak broadening in the related ScZn$_{7.33}$ binary icosahedral quasicrystal \cite{Canfield10,Goldman11} are essentially identical to what we observe here for the i-$R$-Cd family suggesting that the degree of phason strain is endemic to this subclass of the Tsai-type quasicrystals.

\subsection{Magnetization}

\begin{figure}[tbh!]
\includegraphics[scale = 0.3]{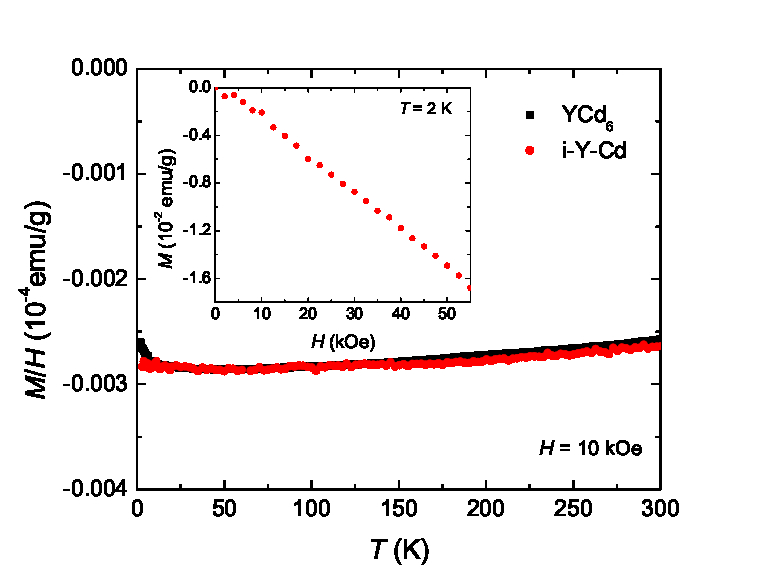}
\caption{(Color online) Temperature-dependent dc magnetic susceptibility of i-Y-Cd and YCd$_{6}$ measured at 10 kOe. The inset shows the field-dependent magnetization of i-Y-Cd measured at 2 K.}
\label{MY}
\end{figure}

The temperature-dependent dc magnetization of i-Y-Cd and YCd$_{6}$ are shown in Fig.~\ref{MY}. Both compounds exhibit diamagnetic and essentially temperature-independent behavior with a value close to -3$\times$10$^{-7}$ emu/g. Compared with other i-$R$-Cd members at room temperature, the absolute value of magnetization for i-Y-Cd is about two orders of magnitude smaller. In addition, the sign and order of magnitude of the dc magnetization is close to another Y-based quasicrystal: Y-Mg-Zn\cite{Fisher99}. At 2 K, the field-dependent magnetization is negative and linear.

\begin{figure}[tbh!]
\includegraphics[scale = 0.3]{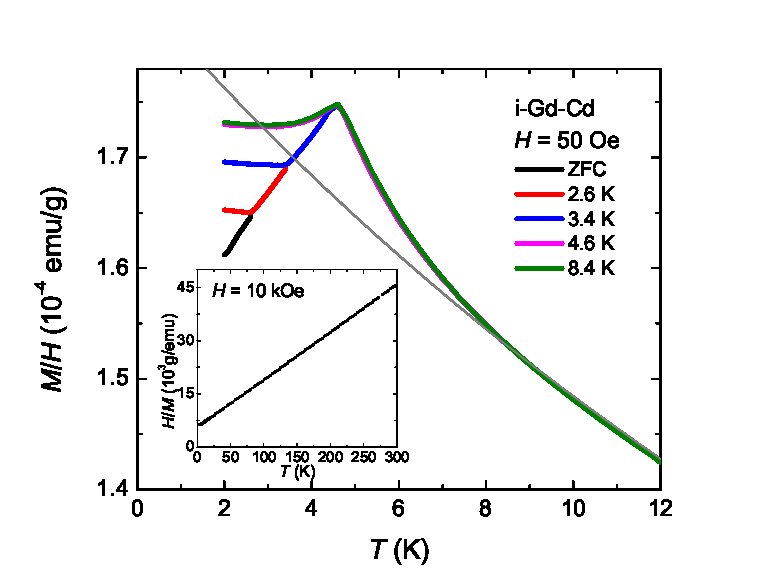}
\caption{(Color online) Irreversibility of dc magnetization measured at 50 Oe. Different colors indicate different field-cool temperatures (see text). Inset shows temperature-dependent inverse magnetic susceptibility of i-Gd-Cd measured at 10 kOe. Grey line represents Curie-Weiss behavior that is extrapolated from its high-temperature paramagnetic state.}
\label{dcGd}
\end{figure}

The inverse magnetic susceptibility for i-Gd-Cd is linear from 300 K down to about 10 K as shown in the inset of Fig.~\ref{dcGd}. i-Gd-Cd exhibits a typical spin glass behavior with a clear cusp at 4.6 K in the zero-field-cooled (ZFC) magnetization data. Below 4.6 K, the field-cooled (FC) magnetization also exhibits a small cusp and then remains almost temperature-independent. In Fig.~\ref{dcGd}, we show that the dc magnetization is strongly history-dependent: data were measured following an initial zero-field cooling to 2 K and warming in a 50 Oe field from 2 K to various temperatures, $T$'. For instance, after ZFC measurement from 2 K to 2.6 K, the sample was cooled with applied field back to 2 K, after which the magnetization data was acquired upon warming from 2 K to 3.4 K. Therefore, the red-line in Fig.~\ref{dcGd} can be considered as a 2.6 K FC measurement. The magnetization after field-cooling from various $T$' are essentially temperature-independent up to $T$' and then fall back onto the ZFC data above $T$'. Here we define $T_\text{irr}$ as the highest temperature where ZFC and FC data differ by more than 0.5\% and $T_\text{max}$ as the temperature at which the maximum of ZFC dc magnetization occurs. In the case of i-Gd-Cd, both characteristic temperatures are the same. Above the $T_\text{max}$, the FC and ZFC data for i-Gd-Cd are essentially identical. In addition, a Curie-Weiss extrapolation from the high-temperature, paramagnetic state of i-Gd-Cd is plotted in grey solid line in Fig.~\ref{dcGd}. There exists a clear deviation from the Curie-Weiss behavior at a higher temperature than $T_\text{max}$. This precursor of spin-glass state may imply a formation of magnetic clusters prior to the spin freezing temperature\cite{Fisher99,Mydosh82,Mydosh83a,Mydosh83b,Binder86}.

\begin{figure}[tbh!]
\includegraphics[scale = 0.3]{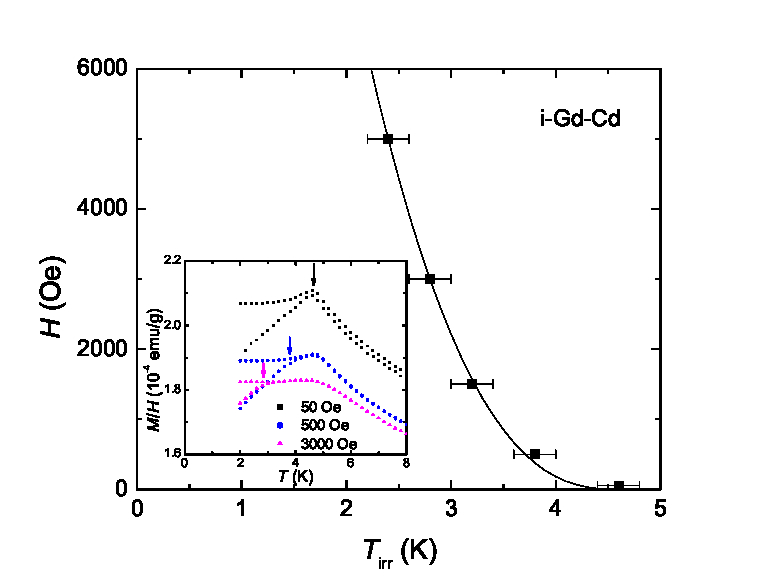}
\caption{(Color online) The field dependence of $T_\text{irr}$ for i-Gd-Cd. Fits according to Eq.~\ref{AT} are shown with $b$ = 2.5 and $\alpha$ = 3.3 $\times$ 10$^{4}$ Oe. The inset shows representative ZFC and FC measurements under different applied magnetic filed. Arrows indicate $T_\text{irr}$ for different fields.}
\label{AT line}
\end{figure}

The dc magnetization of i-Gd-Cd was also investigated at different applied fields and the results are presented in Fig.~\ref{AT line}. The applied field has two significant effects on the measured magnetization: the first being that the cusp in ZFC measurement is rounded and broadened and the second is that $T_\text{irr}$ is shifted to lower temperatures with higher applied fields. A subset of the data is presented in the inset to illustrate these effects. The onset of irreversibility can be associated with de Almeida-Thouless's prediction\cite{Almeida}, where the change in $T_\text{irr}$ with applied field should follow:

\begin{equation}
 H(T_\text{irr}) = \alpha (1 - \frac{T_\text{irr}}{T_\text{f}})^{b}
\label{AT}
\end{equation} 

\noindent
where $T_\text{f}$ stands for the spin freezing temperature in zero field which for i-Gd-Cd was taken as 4.6 K, the same value as $T_\text{max}$. $\alpha$ is the applied magnetic field, above which the irreversibility phenomenon of spin-glass should be fully suppressed. The data for i-Gd-Cd can be fitted with $\alpha$ = 3.3($\pm$ 0.3) $\times$ 10$^{4}$ Oe and $b$ = 2.5 ($\pm$ 0.1). This value of $\alpha$ is close to that found for Tb-Mg-Zn which has an $\alpha$ = 3.5 $\times$ 10$^{4}$ Oe\cite{Fisher99}. It should be noted, though, that the original theory was developed for Ising spins with $b$ = 1.5. Clearly, this will not be the case for Gd moments. For Heisenberg spins, however, an even smaller value of $b$ = 0.5 was predicted\cite{GT81} and could not give a reasonable fit to our data. We also note that, in Ref.~\onlinecite{Fisher99}, the fit could be improved if a larger value of exponent was used. However, it is not clear what causes the difference in the exponent values. 

Compared with a long-range magnetic ordering, which can be viewed as a thermal equilibrium state during the time scale of measurement, a spin-glass is not in such an equilibrium state. Therefore, the magnetic behavior will depend on the frequency of measurement due to the system's limited ability to respond to the changing applied field. The cusp-temperature in the real part of the ac susceptibility increases by about 0.16 K upon increasing the measurement frequency from 10 Hz to 10000 Hz. This implies about 3$\%$ increase of the freezing temperature, which is close to that found for Gd-Mg-Zn\cite{Budko2012}. $\Delta T_\text{f}/[T_\text{f}\Delta(log_{10}f)]$ is about 0.01.

Although the dc magnetization data and the frequency dependence of ac magnetization are consistent with a spin-glass-type freezing of the magnetic moments in i-Gd-Cd, more evidence is required to rule out superparamagnetic-type blocking, or cluster glass, behavior. A convincing way to distinguish between these possibilities is to look at the third order, non-linear magnetic susceptibility, $\chi_{3}$, in the vicinity of spin-freezing/blocking temperature\cite{Bitoh96,mydosh}. $\chi_{3}$ is defined in terms of magnetization, $M$, and applied field, $H$, as follows,

\begin{equation}
M/H = \chi = \chi_{1} + \chi_{3} H^{2} + \chi_{5} H^{4} + \cdots
\end{equation}

\begin{figure}[tbh!]
\includegraphics[scale = 0.3]{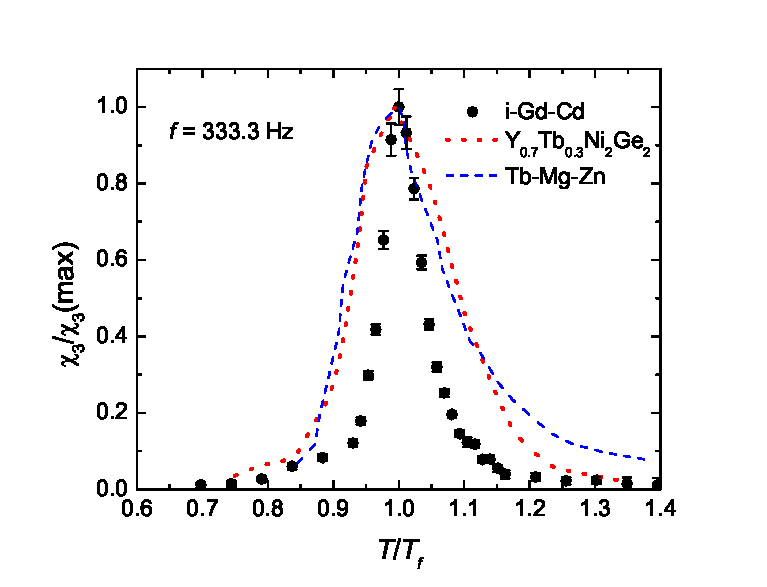}
\caption{(Color online) Third order magnetic susceptibility term, $\chi_{3}$, normalized in temperature and peak height. For i-Gd-Cd, $\chi_{3}$ was measured at 333.3 Hz (black dots). Red and blue dashed lines present the $\chi_{3}$ data, obtained from Ref.~\onlinecite{Wiener00,Fisher99}, for Y$_{0.7}$Tb$_{0.3}$Ni$_{2}$Ge$_{2}$ and Tb-Mg-Zn.}
\label{chi3}
\end{figure}

The temperature-dependent $\chi_{3}$ term will exhibit a much sharper peak in spin-glass systems as compared with a broad feature that is usually observed in superparamagnets\cite{Bitoh96}. The third order magnetic susceptibility was investigated for i-Gd-Cd and is shown in Fig.~\ref{chi3}. An ac field with an amplitude of 3 Oe and a frequency of 333.3 Hz was applied to acquire the data. The $\chi_{3}$ peak for i-Gd-Cd is sharper than other known spin-glass systems, for example, Tb-Mg-Zn, Ho-Mg-Zn quasicrystals\cite{Fisher99} and an Ising spin glass system Y$_{1-x}$Tb$_{x}$Ni$_{2}$Ge$_{2}$\cite{Wiener00}, whereas superparamagnets usually exhibit a much broader feature\cite{Bitoh96}. We point out that the peak temperature in $\chi_{3}$ for i-Gd-Cd is 4.3 K, a value that is 0.3 K lower than the $T_\text{max}$ value from the dc magnetization. This small discrepancy in temperatures possibly results from a different thermometry configuration in the QD PPMS where ac susceptibility was measured. For the current study, we follow the temperatures given by MPMS.

\begin{figure}[tbh!]
\includegraphics[scale = 0.32]{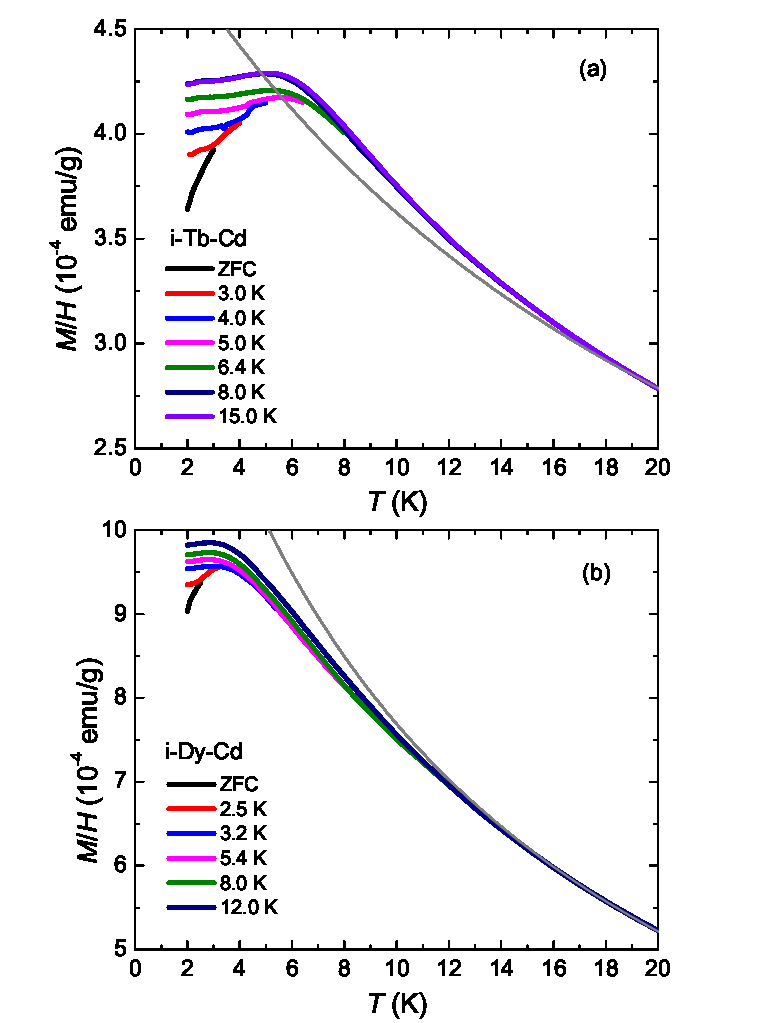}
\caption{(Color online) Temperature-dependent dc magnetization measured at 50 Oe for i-$R$-Cd ($R$ = Tb, Dy). Different colors indicate different FC temperatures. Grey lines represent Curie-Weiss behavior that was extrapolated from the high-temperature paramagnetic state.}
\label{dcTbDy}
\end{figure}

Whereas the data just presented for i-Gd-Cd is consistent with classic spin-glass behavior, for i-Tb-Cd and i-Dy-Cd, the canonical cusp-shaped spin-glass signature in magnetization data is replaced by a broad maximum in both ZFC and FC data with the irreversibility appearing at a significantly higher temperature (Fig.~\ref{dcTbDy}). Similar behavior was observed in the $R$-Mg-Cd quasicrystal system and explained by the presence of magnetic impurities that due to slight oxidation of the surface of the sample\cite{Sebastian}. After annealing at 200$^{\circ}$C for two days, the dc magnetization data for both i-Tb-Cd and i-Ho-Cd remain the same, even though a thin layer of oxidation appeared on the sample's surface. If the aforementioned argument is applied, a magnetization feature that evolves with changing degrees of oxidation would be expected. Sample inhomogeneity that causes non-cusp like feature in other spin glass systems\cite{Chamberlin}, if exists in i-Tb-Cd and i-Dy-Cd, cannot be removed by annealing at 200$^{\circ}$C. 

In Fig.~\ref{dcTbDy}, a clear history-dependent magnetization can be observed. Comparing with i-Gd-Cd, in which different FC temperatures result in a temperature-independent magnetization from the base temperature up to $T$', in the case of i-Tb-Cd and i-Dy-Cd, if $T$' is higher than $T_\text{max}$, the temperature-independence survives only up to $T_\text{max}$. This may indicate that only at temperatures that are lower than $T_\text{max}$, do the magnetic moments become fully "frozen". Therefore, $T_\text{max}$ might represent the spin-freezing temperature, $T_\text{f}$, better than $T_\text{irr}$.

The grey, solid curves shown in Fig.~\ref{dcTbDy} are the extrapolations of the high-temperature Curie-Weiss fit to the data. Comparing with i-Gd-Cd, i-Tb-Cd and other members, the manner in which i-Dy-Cd deviates from the Curie-Weiss behavior is different, since its magnetization increases more slowly upon cooling than its high-temperature, paramagnetic state would suggest. Since i-Gd-Cd, i-Tb-Cd and i-Dy-Cd each exhibit deviation from simple Curie-Weiss behavior, it is clear that the CEF splitting is not the key factor for the formation of possible magnetic clusters. However, subtleties in the CEF splitting might alter the details of magnetic properties and make i-Dy-Cd behave differently. A similar change in the sign of the deviation was also reported in other spin-glass systems, like AuFe alloys, with different Fe concentrations\cite{Mydosh83a}. However, in that case, the sign of Curie-Weiss temperature changes at the same time. 

Attempts to obtain temperature-dependent $\chi_{3}$ data for i-Tb-Cd were made. However, no resolvable feature was detected. Given that no clear cusp was seen in the ZFC dc magnetization measurement, it is likely that, for this compound, a possible distribution of freezing temperatures makes it difficult to experimentally see the clear feature in $\chi_{3}$. However, the experimental limitations of our instruments can not be ruled out.

\begin{figure}[tbh!]
\includegraphics[scale = 0.32]{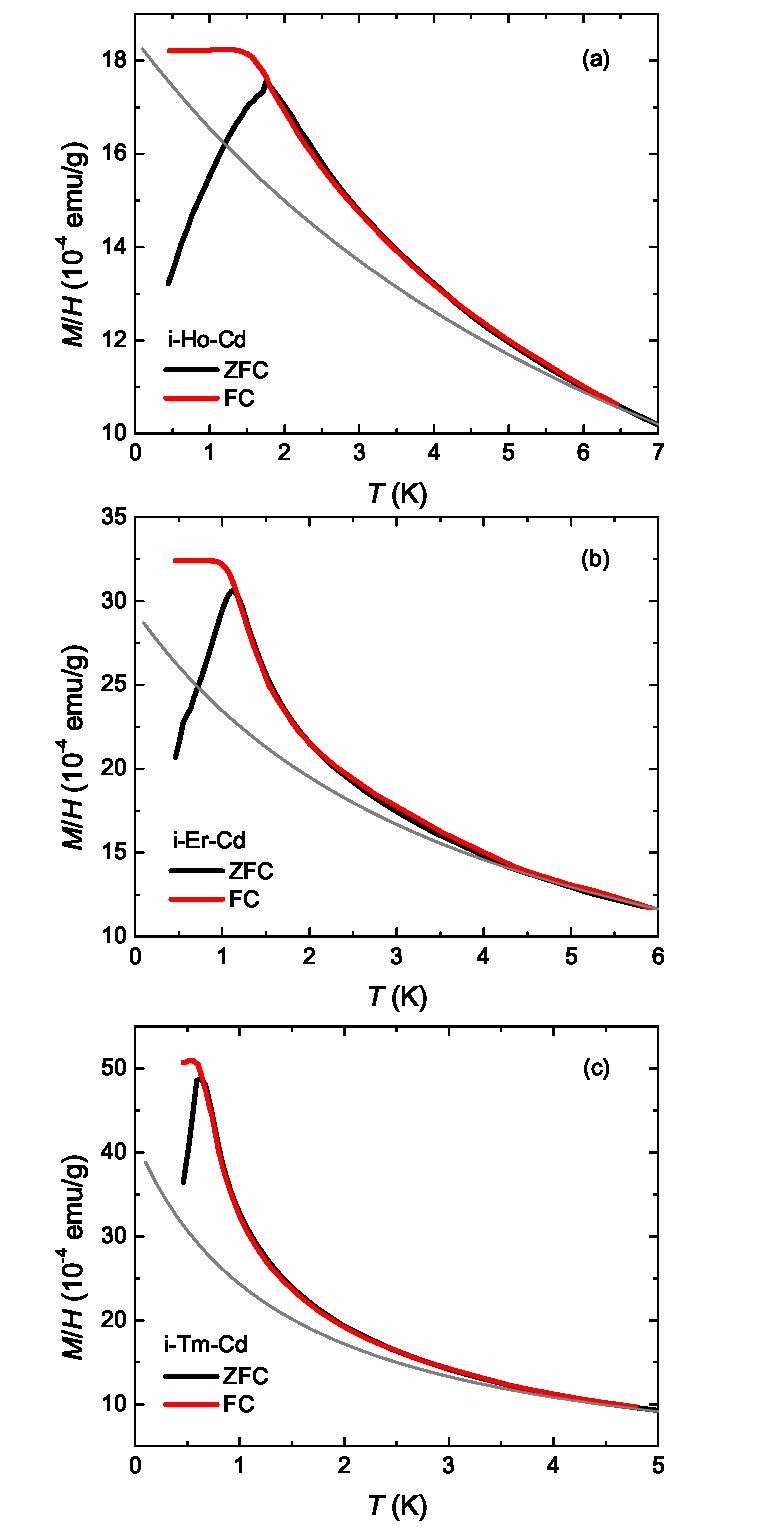}
\caption{(Color online) Temperature-dependent dc magnetization measured at 50 Oe for i-$R$-Cd ($R$ = Ho-Tm). Data were acquired using a QD iHelium3 system. Grey lines represent Curie-Weiss behavior that is extrapolated from the high-temperature paramagnetic state.}
\label{dcHoErTm}
\end{figure}

The dc magnetization data measured in 50 Oe, down to 0.46 K are shown for i-Ho-Cd, i-Er-Cd and i-Tm-Cd in Fig.~\ref{dcHoErTm}. In the ZFC data, clear cusps can be observed with $T_\text{max}$ = $T_\text{irr}$. In general, the irreversibility features for these members are much closer to what was seen in i-Gd-Cd, i.e. sharp cusps in ZFC dc magnetization. One subtle difference being that, unlike the case of i-Gd-Cd where ZFC and FC magnetization data reaches the maximum at the same temperature, for i-$R$-Cd ($R$ = Ho-Tm), the FC maximum is located at a slightly lower temperature than $T_\text{max}$. The deviation from the Curie-Weiss, paramagnetic state (shown in grey) occurs at higher temperature than $T_\text{max}$, with a clear upward deviation. Information obtained from all of our dc magnetization measurements are summarized in Table~\ref{table}.

Due to the limitations of our instrument, we were not able to measure ac magnetization below 1.8 K, where the dc magnetization features of i-$R$-Cd ($R$ = Ho-Tm) emerge.

\begin{table}[h!]
\caption{Characteristic temperatures of i-$R$-Cd ($R$ = Gd-Tm). Curie-Weiss temperatures, $\Theta$, were adopted from Ref.~\onlinecite{QC2013}. $T_\text{max}$ represents the temperature at which dc ZFC data reaches maximum. $T_\text{irr}$ represents the temperature at which FC and ZFC data start to split by more than 0.5$\%$. Error bars were estimated according to measurements on different samples and the data step width of each measurement.}
\begin{tabular}{p{2cm} p{1.0cm} p{1.6cm} p{1.6cm} p{1.4cm}}
\hline
\hline
Compound & $\Theta$ (K)& $T_\text{max}$ (K)& $T_\text{irr}$ (K)\\
\hline
i-Gd-Cd& -41(1)& 4.6(0.2) & 4.6(0.2)\\
i-Tb-Cd& -21(1)& 5.3(0.5) & 8.7(0.5)\\
i-Dy-Cd& -11(1)& 3.0(0.4) & 10.1(0.3)\\
i-Ho-Cd& -6(1)& 1.76(0.05) & 1.76(0.05)\\
i-Er-Cd& -4(1)& 1.11(0.05) & 1.11(0.05)\\
i-Tm-Cd& -2(1)& 0.63(0.05) & 0.63(0.05)\\
\hline
\end{tabular}
\label{table}
\end{table}

\subsection{Specific Heat}

\begin{figure}[tbh!]
\includegraphics[scale = 0.3]{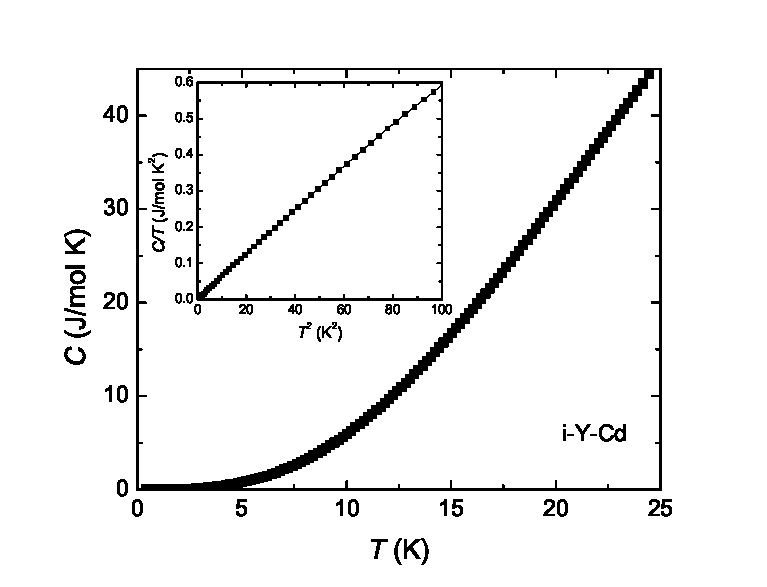}
\caption{Temperature-dependent specific heat for i-Y-Cd using a stoichiometry of YCd$_{7.48}$. The inset shows C$_{p}$/$T$ versus $T^{2}$ up to 10 K}
\label{CY}
\end{figure}

The temperature-dependent specific heat of i-Y-Cd is shown in Fig.~\ref{CY}. The stoichiometry used for the calculation was adopted from the Wavelength Dispersive Spectroscopy (WDS) results reported in Ref.\onlinecite{QC2013}, YCd$_{7.48}$ for i-Y-Cd in this case. Below 10 K, there is a linear region in $C$/$T$ versus $T^{2}$, which yields a Debye temperature, $\Theta_\text{D}$, of about 140 K. The linear fit also intersects the $C$/$T$ axis at roughly 4 $\pm$ 2 mJ/mol-Y K$^{2}$ (or 0.5 mJ/atom K$^{2}$). For crystalline solids, the intercept normally indicates the electronic specific heat, $\gamma$. However, it was also noticed that even for non-crystalline solids, there could still be a linear region in $C$/$T$ versus $T^{2}$ plot at low temperature with a finite $\gamma$ value\cite{Talon02}, which can be explained by a distribution of two-level systems\cite{Phillips87}. If this is the case, then the electronic specific heat contribution to $C$/$T$ will be even closer to zero. If we take the measured value, 4 $\pm$ 2 mJ/mol-Y K$^{2}$, as the $\gamma$ for i-Y-Cd, both $\gamma$ and $\Theta_{D}$ for i-Y-Cd are very close to the values obtained for YCd$_{6}$\cite{Mori12}. 
 
\begin{figure}[tbh!]
\includegraphics[scale = 0.32]{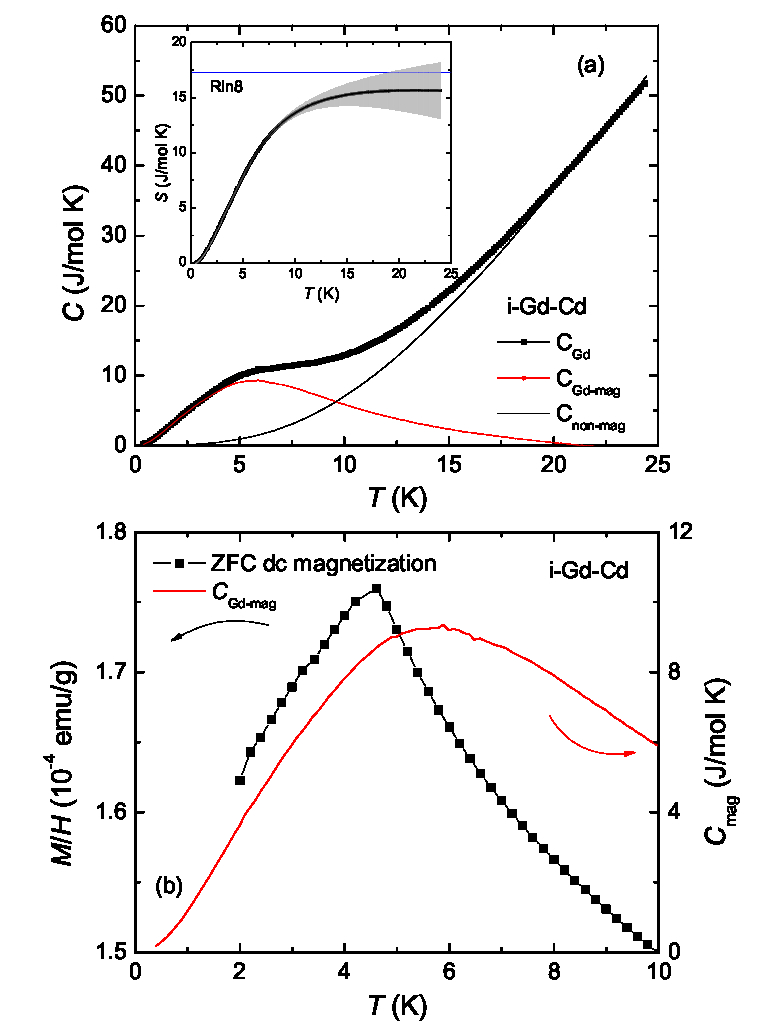}
\caption{(Color online) (a) Temperature-dependent specific heat of i-Gd-Cd. The grey solid line represents the non-magnetic part of the specific heat. The red solid line shows the magnetic specific heat. The inset shows the magnetic entropy with grey error bars (see text). (b) Low-temperature magnetic specific heat (red) on the right scale and low-temperature ZFC dc magnetization (black) on the left scale.}
\label{Gd}
\end{figure}

\begin{figure}[tbh!]
\includegraphics[scale = 0.32]{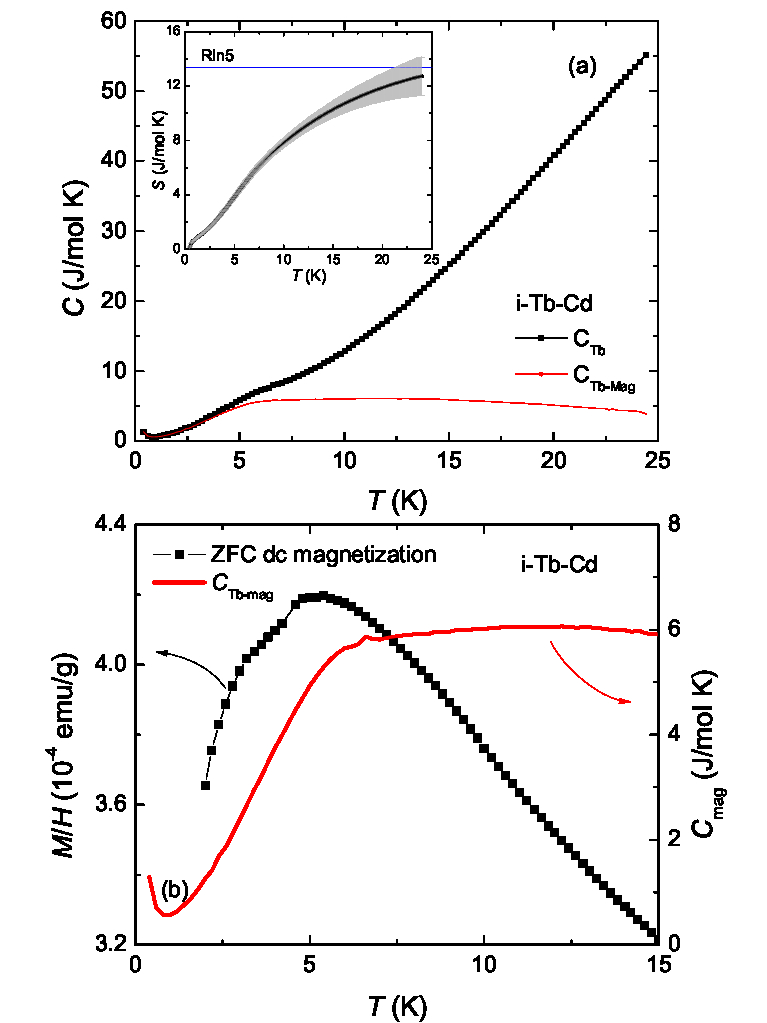}
\caption{(Color online) (a) Temperature-dependent specific heat of i-Tb-Cd. The red solid line shows the magnetic specific heat. The inset show the magnetic entropy with error bars. (b) Low-temperature magnetic specific heat (red) on the right scale and low-temperature ZFC dc magnetization (black) on the left scale.}
\label{Tb}
\end{figure}

\begin{figure}[tbh!]
\includegraphics[scale = 0.32]{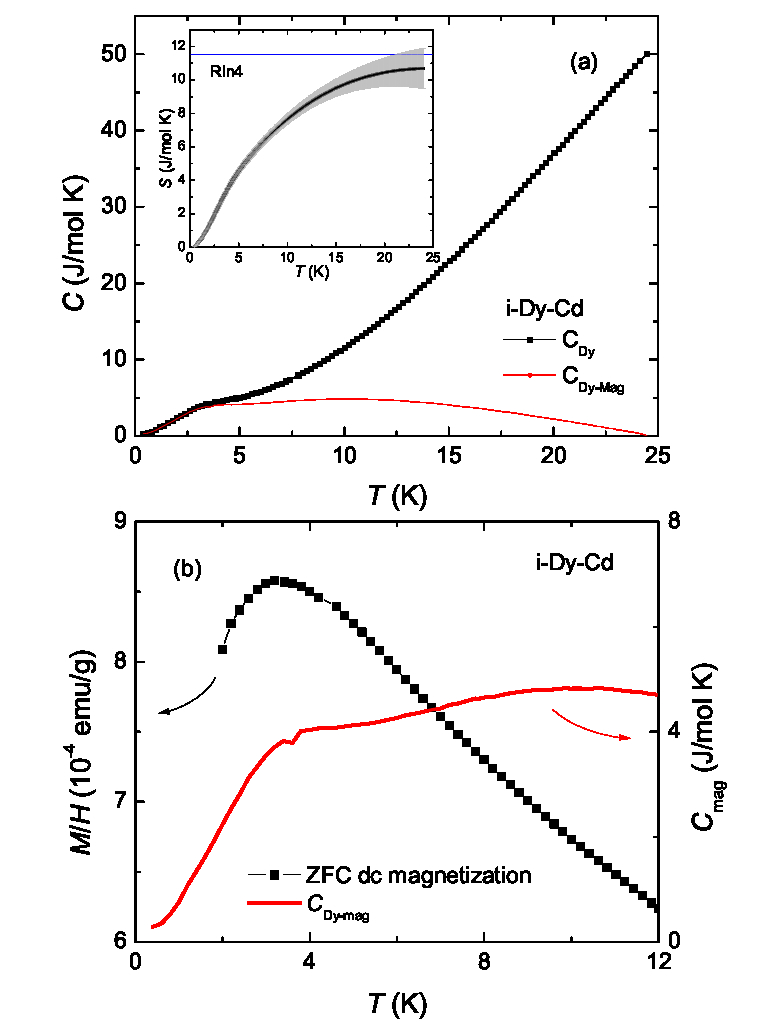}
\caption{(Color online) (a) Temperature-dependent specific heat of i-Dy-Cd. The red solid line shows the magnetic specific heat. The inset show the magnetic entropy with error bars. (b) Low-temperature magnetic specific heat (red) on the right scale and low-temperature ZFC dc magnetization (black) on the left scale.}
\label{Dy}
\end{figure}

\begin{figure}[tbh!]
\includegraphics[scale = 0.32]{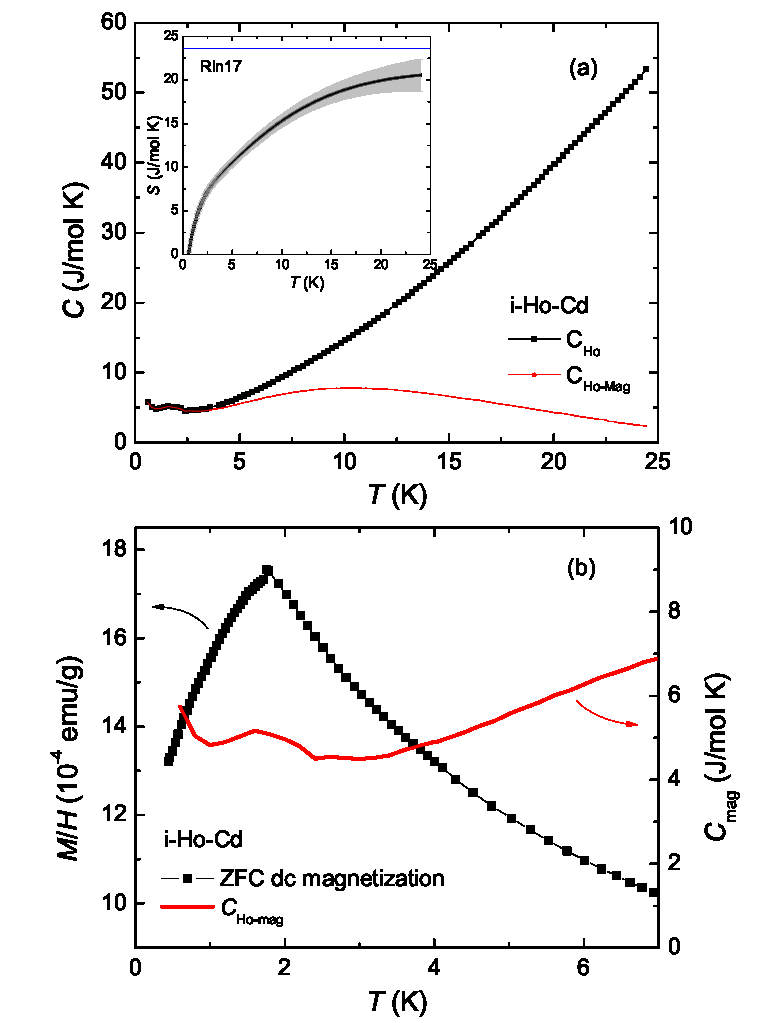}
\caption{(Color online) (a) Temperature-dependent specific heat of i-Ho-Cd. The red solid line shows the magnetic specific heat. The inset shows the magnetic entropy with error bars. (b) Low-temperature magnetic specific heat (red) on the right scale and low-temperature ZFC dc magnetization (black) on the left scale.}
\label{Ho}
\end{figure}

\begin{figure}[tbh!]
\includegraphics[scale = 0.32]{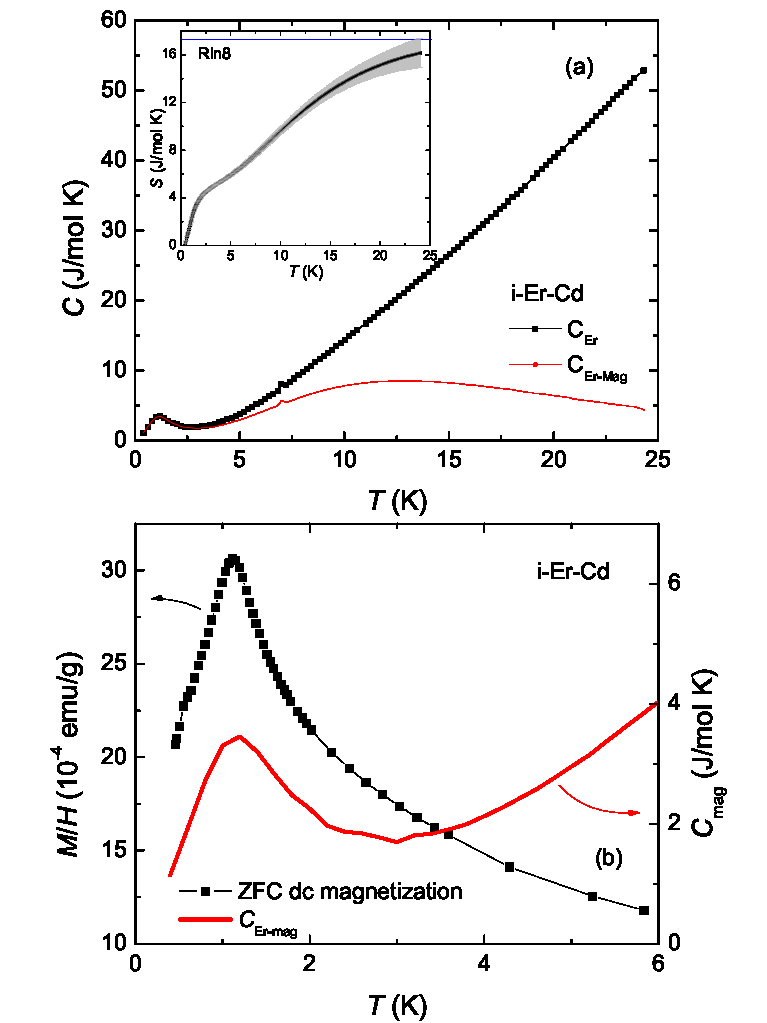}
\caption{(Color online) (a) Temperature-dependent specific heat of i-Er-Cd. The red solid line shows the magnetic specific heat. The inset shows the magnetic entropy with error bars. (b) Low-temperature magnetic specific heat (red) on the right scale and low-temperature ZFC dc magnetization (black) on the left scale.}
\label{Er}
\end{figure}

\begin{figure}[tbh!]
\includegraphics[scale = 0.32]{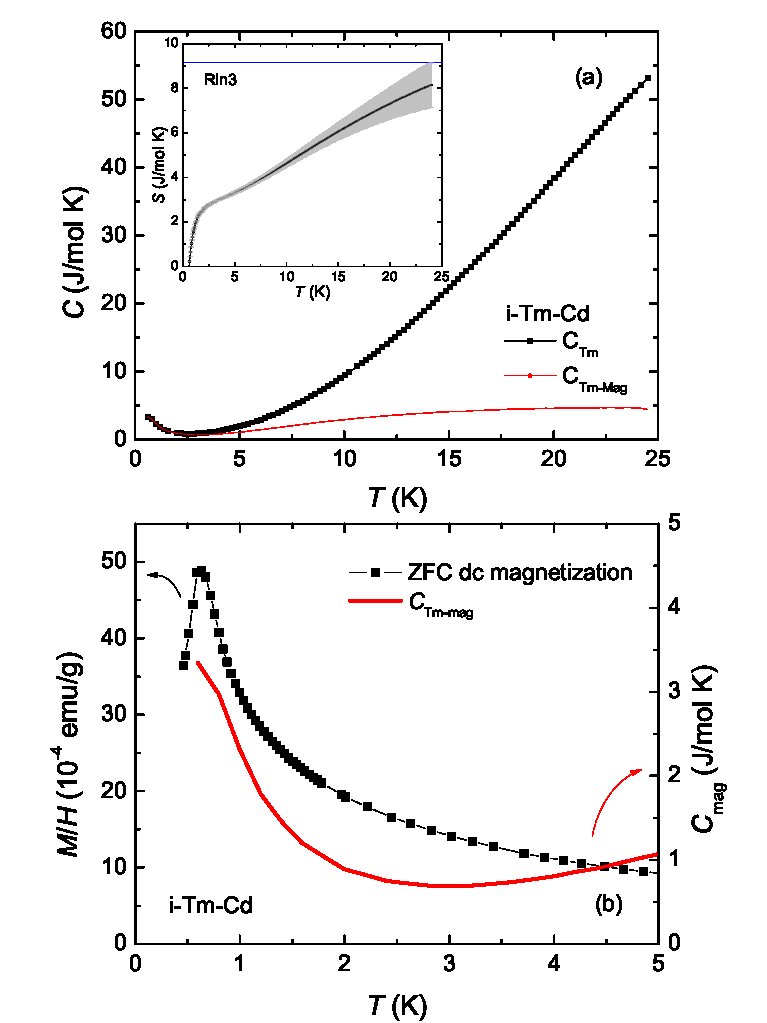}
\caption{(Color online) (a) Temperature-dependent specific heat of i-Tm-Cd. The red solid line shows the magnetic specific heat. The inset shows the magnetic entropy with error bars. (b) Low-temperature magnetic specific heat (red) on the right scale and low-temperature ZFC dc magnetization (black) on the left scale.}
\label{Tm}
\end{figure}

Figs.~\ref{Gd}-\ref{Tm} present the specific heat data for other i-$R$-Cd members. The magnetic specific heat was calculated by subtracting that of i-Y-Cd with a small molar mass corrections according to (1) the Debye model to approximate changes due to the heavier $R$ ions and (2) the $R$ concentration. Although it is not well investigated if the model works for quasicrystalline compounds, the magnetic entropy thus integrated offers some information about how CEF splitting of the Hund's rule ground state multiplet $J$ of the $R^{3+}$ ion plays a role in the magnetism, as well as the temperatures at which the magnetic entropy starts to change. Uncertainties, shown as grey areas in Figs.~\ref{Gd}-\ref{Tm}, take into account the uncertainty in the WDS-determined stoichiometry and the uncertainty in the mass of the sample. Any un-physical drop in magnetic entropy at high temperature, due to this increasing error bar, can be ignored. 

In Fig.~\ref{Gd}, i-Gd-Cd shows a typical spin-glass behavior with a broad maximum located roughly 20$\%$ above $T_\text{f}$\cite{Binder86, mydosh}. Above roughly 10 K, the magnetic entropy of i-Gd-Cd tends to saturate reaching the expected value for non-CEF-split Gd$^{3+}$, $\mathcal{R}$ln8, where $\mathcal{R}$ is the universal gas constant. The temperature where the magnetic entropy of i-Gd-Cd starts to saturate roughly corresponds to the temperature at which precursor magnetic clusters start to form as inferred from the deviation from the high-temperature Curie-Weiss tail seen in the dc magnetization measurements (see Fig.~\ref{dcGd}). 

For the rest of the i-$R$-Cd members, the CEF splitting, albeit relatively small in this high symmetry structure\cite{Walter87}, lifts the degeneracy of trivalent rare earth ground state. This results in a slower recovery of the full $\mathcal{R}$ln(2$J$+1) magnetic entropy upon warming and the thermal excitations between split levels persist to higher temperature as compared with the case of i-Gd-Cd. 

The specific heat of i-Tb-Cd and i-Dy-Cd, members that exhibit non-cusp-like features in low-field magnetization data, are shown in Figs.~\ref{Tb}-\ref{Dy}. In the specific heat of i-Tb-Cd, the broad peak is not as clear as in i-Gd-Cd, which is possibly due to the addition of Schottky anomalies to the background. The origin of the slight low-temperature upturn observed below 1 K, however, is not yet well understood. A similar upturn is also observed in i-Ho-Cd and presented in Fig.~\ref{Ho} below. Since among all the studied i-$R$-Cd members, $R$ = Tb and Ho have the largest gyromagnetic ratio for the nuclear spins, it is likely that this low-temperature upturn in the specific heat originates from a nuclear Schottky anomaly\cite{NMR1}. If the low-temperature specific heat upturn is included, the magnetic entropy of i-Tb-Cd is about $\mathcal{R}$ln2 at $T_\text{max}$ and $\mathcal{R}$ln4 at the temperature where the dc magnetization starts to deviate from the Curie-Weiss behavior. Lacking more low-temperature data for the Schottky anomaly fit, it is difficult to offer quantitative corrections to the magnetic entropy. Qualitatively, the magnetic entropy for i-Tb-Cd may decrease by about 1 J/mol K if the upturn feature is excluded.

The specific heat of i-Dy-Cd is similar with that of i-Tb-Cd. The magnetic entropy reaches $\mathcal{R}$ln2 at around 5 K and approaches $\mathcal{R}$ln4 at 24 K. The magnetic specific heat shown in solid red line exhibits a broad maximum in the vicinity of the broad feature observed in the ZFC magnetization data. Another broad hump centered at $\sim$10 K is most likely associated with Schottky anomalies due to CEF split levels.

Apart from the upturn at low temperatures, the specific heat of i-Ho-Cd (shown in Fig.~\ref{Ho}) is different in a sense that it recovers the magnetic entropy much faster. After a subtraction by i-Y-Cd, a large amount of magnetic contribution in specific heat still exists below 5 K. At 24 K, the calculated magnetic entropy approaches $\mathcal{R}$ln17, which is the full magnetic entropy expected for Ho$^{3+}$. Even if the specific heat upturn below 1 K is assumed to arise from a nuclear Schottky anomaly, and is therefore excluded, an uncertainty of up to 4 J/mol K still suggests an $\mathcal{R}$ln9 magnetic entropy at 24 K. This large amount of entropy implies relatively small CEF splitting and is consistent with a distribution of low laying Schottky anomalies. In addition, a small, rounded, hump next to $T_\text{max}$ might be consistent with a spin-glass transition.

The specific heat of i-Er-Cd and i-Tm-Cd are shown in Figs.~\ref{Er}-\ref{Tm}. i-Er-Cd exhibits a hump in specific heat. At 24 K, it approaches $\mathcal{R}$ln8. The magnetic specific heat, in addition, shows another clear broad hump at around 12 K. This is most likely due to a Schottky anomaly associated with undetermined CEF levels. However, the current data does not allow for a more detailed analysis.

In i-Tm-Cd, we only observed part of the specific hump due to our base temperature of PPMS. It should be noted that it seems that the maximum temperature in specific heat of i-Tm-Cd is equal to, or even lower than the $T_\text{max}$ in the dc magnetization data. 

\section{Discussion}

The existence of i-$R$-Cd ($R$ = Y, Gd-Tm) allows for the study of systematic trends across this binary, local-moment-bearing quasicrystal series. As shown in Fig.~\ref{xray}, there is a standard lanthanide contraction associated with changing $R$. Despite possible slight change in stoichiometry, there is no clear difference between i-Gd-Cd and i-Tm-Cd in terms of strain and phason strain, or in other words, sample quality.

For i-Gd-Cd, we have presented data that support the identification of $T_\text{max}$ as $T_\text{f}$, the spin glass freezing temperature. These data include: (1) a cusp in dc magnetization; (2) a frequency-dependent freezing temperature; (3) a narrow third order magnetic susceptibility, $\chi_{3}$, at the freezing temperature; (4) a broad maximum in temperature-dependent specific heat with the maxima temperature somewhat higher than the cusp temperature in dc magnetization. According to the general understanding of the experimental characteristics of spin-glasses\cite{mydosh}, i-Gd-Cd can be considered to be a spin-glass below $T_\text{f}$ = 4.6 K. Unfortunately, we were unable to obtain as extensive sets of data for the rest of the i-$R$-Cd ($R$ = Tb-Tm) series, especially $\chi_{3}$. In discussion of the magnetization features, i-Tb-Cd and i-Dy-Cd are of special interest for their non-cusp-like ZFC dc magnetization data. However, in view of the similarity in resistance (see Appendix) and specific heat data, it is likely that neither i-Tb-Cd nor i-Dy-Cd exhibit long-range magnetic ordering. Further investigations of i-Tb-Cd and i-Dy-Cd are needed to elucidate the origin of this non-standard spin-glass-like behavior in the magnetization. The rest of the local-moment-bearing members, i-$R$-Cd ($R$ = Ho-Tm), behave closer to a canonical spin-glass in terms of their dc magnetization. To obtain $\chi_{3}$ for these three members, ac magnetization measurements below 2 K are needed, which is currently beyond our instrumental capability. It is worth pointing out that although broadened maxima, rather than $\lambda$-like peaks, were observed in the specific heat measurements, the maximum temperatures are close to, if not equal to, $T_\text{max}$. Whereas in canonical spin-glass systems, the broad maximum in specific heat often exceeds the freezing temperature, $T_\text{f}$, by about 20\%-50\%\cite{Binder86, mydosh}. 

\begin{figure}[tbh!]
\includegraphics[scale = 0.3]{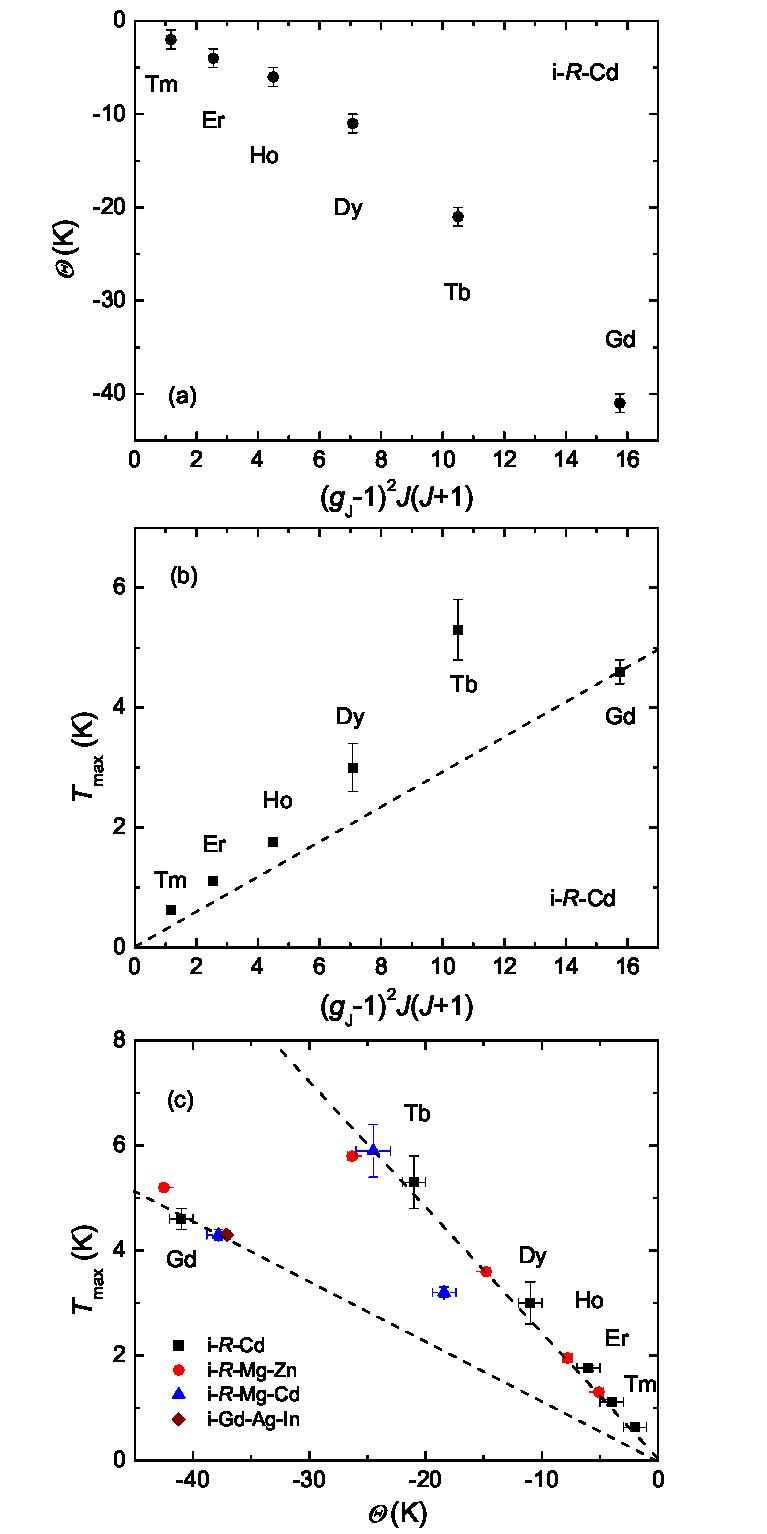}
\caption{(Color online) (a)/(b) Changes of $\Theta$/$T_{max}$ as a function of de Gennes factor: ($g_{J}$-1)$^{2}J$($J$+1). (c) Changes of $T_\text{max}$ as a function of $\Theta$. Data for i-$R$-Mg-Zn, i-$R$-Mg-Cd and i-Gd-Ag-In are obtained from Refs.~\onlinecite{Fisher99,Canfield01b,Sebastian,stadnik07}. Dashed lines are guides to the eyes.}
\label{dG}
\end{figure}
 
In a series of iso-structural rare earth based compounds, systematic trends in the physical properties are generally expected. According to the de Gennes scaling, the Curie-Weiss temperatures, $\Theta$, are suppose to scale linearly with de Gennes factor dG = ($g_{J}$-1)$^{2}J$($J$+1). In Fig.~\ref{dG}(a), $\Theta$ for i-$R$-Cd ($R$ = Gd-Tm) are plotted against dG. There is a rough agreement between the de Gennes scaling and the experimental data for i-$R$-Cd, similar to that found for different rare-earth-bearing quasicrystal systems\cite{QC2013}.

$T_\text{max}$ is plotted against dG factor and experimentally measured $\Theta$ in Figs.~\ref{dG}(b) and (c). Both show non-monotonic behavior. The clear deviation from de Gennes scaling is evidenced by a higher $T_\text{max}$ of i-Tb-Cd and i-Dy-Cd in Fig.~\ref{dG}(b). In magnetically ordered systems, it was argued that CEF effects can enhance the ordering temperature in materials with a strong axial anisotropy\cite{Noakes82,Dunlap84}. In quasicrystals, although anisotropy is not well defined, the CEF effects do exist for rare earth ions that have a finite orbital angular momentum of their 4$f$ electrons. It might be possible that a small CEF effect\cite{Walter87} helps to stabilize the freezing in this geometrically frustrated system. This is consistent with Fig.~\ref{dG}(c) that shows i-Gd-Cd has a lower $T_\text{max}$ for given $\Theta$. Fig.~\ref{dG}(c) shows a clear difference in $T_\text{max}$/$\Theta$ between Gd$^{3+}$ and the rest of the members as indicated by the dashed lines. In Ref.~\onlinecite{Canfield01b}, this phenomena was associated with the difference between Heisenberg like ion (Gd$^{3+}$) and non-Heisenberg like ion (Tb$^{3+}$-Tm$^{3+}$). It is worth pointing out that, in addition to the similarity in Curie-Weiss temperatures\cite{QC2013}, the value for $T_\text{max}$/$\Theta$ is also similar for different rare earth bearing quasicrystal systems\cite{Fisher99,Canfield01b,Sebastian,stadnik07}. In the plot of $T_\text{max}$/$\Theta$, the slope is $\sim$0.11 for Heisenberg-like ion and $\sim$0.25 for non-Heisenberg-like ions. Both numbers indicate a moderate degree of geometrical frustration\cite{ramirez94}.

\section{Conclusions}

In this study, we presented detailed structural, thermodynamic and transport measurements on i-$R$-Cd ($R$ = Y, Gd-Tm) grown via the solution growth method. Structurally, the clear trend of diffraction peaks broadening as a function of $G_{\rm{\bot}}$ indicates that frozen-in phason strain is the key mechanism for structural disorder in these quasicrystallne samples. No significant difference exists in strain/phason strain between i-Gd-Tm and i-Tm-Cd. 

The magnetic susceptibility of i-Y-Cd is essentially temperature-independent and weakly diamagnetic. The low-temperature specific heat of i-Y-Cd reveals a Debye temperature of about 140 K. Supported by the magnetization and specific heat data, i-Gd-Cd can be categorized as a spin-glass below $T_\text{f}$ = $T_\text{max}$ = 4.6 K. The dc magnetization data of i-Tb-Cd and i-Dy-Cd do not show a typical cusp-like shape but rather a broad feature with a clear temperature spacing between $T_\text{max}$ and $T_\text{irr}$. Further study is needed to explain this unconventional behavior. However, based on the similarity of temperature-dependent resistance and specific heat measurements, it is unlikely that i-Tb-Cd or i-Dy-Cd exhibits long-range magnetic ordering. i-$R$-Cd ($R$ = Ho-Tm) show conventional spin-glass behavior in their magnetization, but with the maximum in the magnetic component of specific heat occurring at temperatures closer and closer to $T_\text{max}$. Further investigation is needed to explain this trend. A deviation from the de Gennes scaling for the moment-bearing members was observed. It is likely this deviation is a consequence of CEF effects, which helps  to stabilize the freezing state of magnetic rare earth ions with finite orbital angular momentum. Resemblance was also noted in the value of $T_\text{f}$/$\Theta$ between i-$R$-Cd and another rare earth based quasicrystal systems.

\section*{Acknowledgements}
We would like to thank A. Sapkota, M. Ramazanoglu, and D. Robinson for the help with the high-energy x-ray diffraction measurement. This work was supported by the U.S. Department of Energy (DOE), Office of Science, Basic Energy Sciences, Materials Science and Engineering Division. The research was performed at the Ames Laboratory, which is operated for the U.S. DOE by Iowa State University under contract NO. DE-AC02-07CH11358.

\section*{Appendix}

Due to a low density of states at the Fermi level, quasicrystals are generally bad metals or sometimes on the edge of a metal-insulator transition\cite{Janot1997,Nayak12}. The resistivity of quasicrystalline material is nearly temperature-independent, or decreases weakly with increasing temperature. Because of the high resistivity of quasicrystals, a small amount of conducting impurity in/on the sample is sufficient to result in significant changes in measured resistance. 

\begin{figure}[tbh!]
\includegraphics[scale = 0.32]{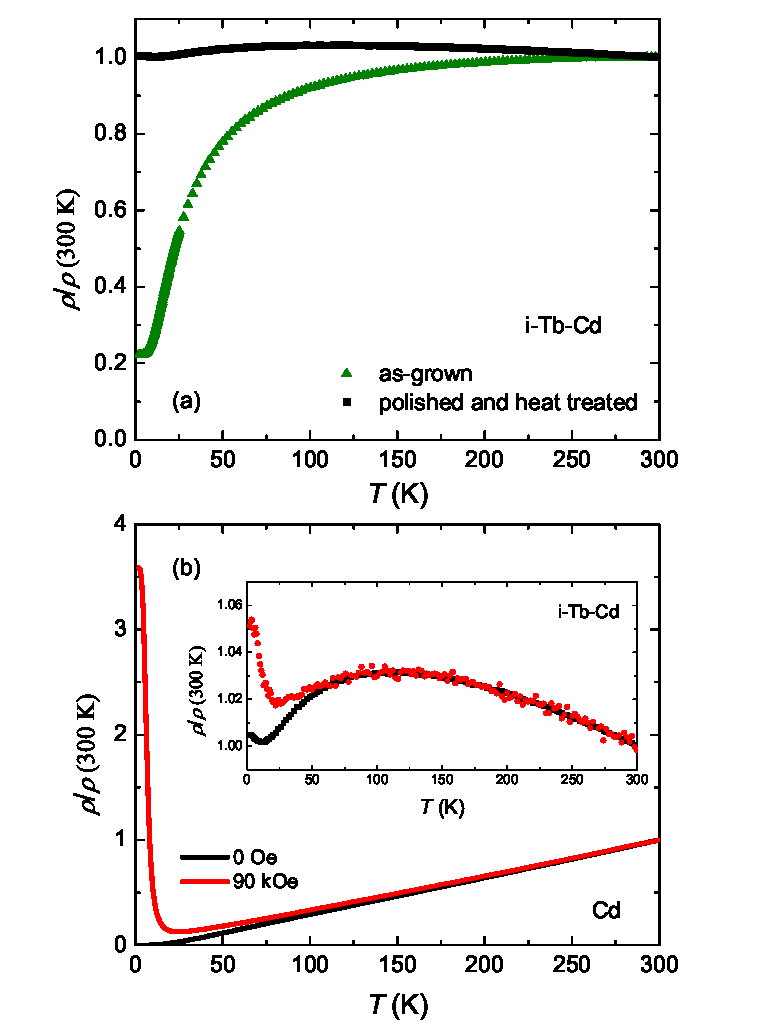}
\caption{(Color online) (a) Zero-field, normalized temperature-dependent resistance for i-Tb-Cd: as-grown sample (green triangle); polished and heat treated sample (black square). (b) Normalized temperature-dependent resistance of elemental Cd (polished and heat treated i-Tb-Cd sample in the inset) measured at zero field (black) and at 90 kOe (red).}
\label{Cd}
\end{figure}

In our initial attempts to measure the electric resistance of i-$R$-Cd, a standard four-probe technique was used and Pt wires were attached to an as-grown sample like shown in Fig.~\ref{diagram}(b). In Fig.~\ref{Cd}(a), the temperature-dependent resistance of as-grown i-Tb-Cd is presented (green triangles). A sizeable residual resistance ratio (RRR) of about 4.5 was obtained, which is distinct from known quasicrystal behavior\cite{Janot1997}. However, the shape of the observed resistance can result from measuring a nearly temperature-independent resistor (the quasicrystal sample) connected in series and in parallel with a highly conducting metal, in this case, Cd. 

We tried to remove the residual Cd in the following way. The sample is sealed in a long quartz tube under vacuum, in which the sample is held at 200$^{\circ}$C while the other end of the tube is held at room temperature. Due to its high vapour pressure, Cd can be easily removed from the sample by this heat treatment. In preparing the sample for resistance measurements, polishing as the first step could remove residue surface Cd and the heat treatment, as the second step, could remove part of the remaining Cd that was trapped in exposed dendritic grain boundaries. After the polished sample went through the heat treatment for 3 days, a nearly temperature-independent resistance was indeed observed, as shown in Fig.~\ref{Cd}(a) by black squares. However, as illustrated by the inset of Fig.~\ref{Cd}(b), the resistance of i-Tb-Cd still decreases below about 100 K, and a clear magnetic field dependence of this low-temperature resistance emerges. For clarity, the resistance of elemental Cd was also measured. The resistance sample of Cd was prepared by pressing an elemental Cd droplet to reduce the thickness. A RRR of about 2200 was observed [Fig.~\ref{Cd}(b)]. The magnetoresistance increases at low temperatures as expected in general for a simple, high purity, metal, and follows Kohler's rule\cite{Abrikosov88,Budko98}. The resemblance of the low-temperature magnetoresistance of i-Tb-Cd to that of Cd suggests that even after polishing and the heat treatment, Cd that is trapped within the qusicrystal might still affect the resistance data. Therefore, it is likely that, despite efforts to eliminate Cd, the features observed are a combination of intrinsic quasicrystal properties plus a minor amount of conducting metal.

\begin{figure}[tbht!]
\includegraphics[scale = 0.3]{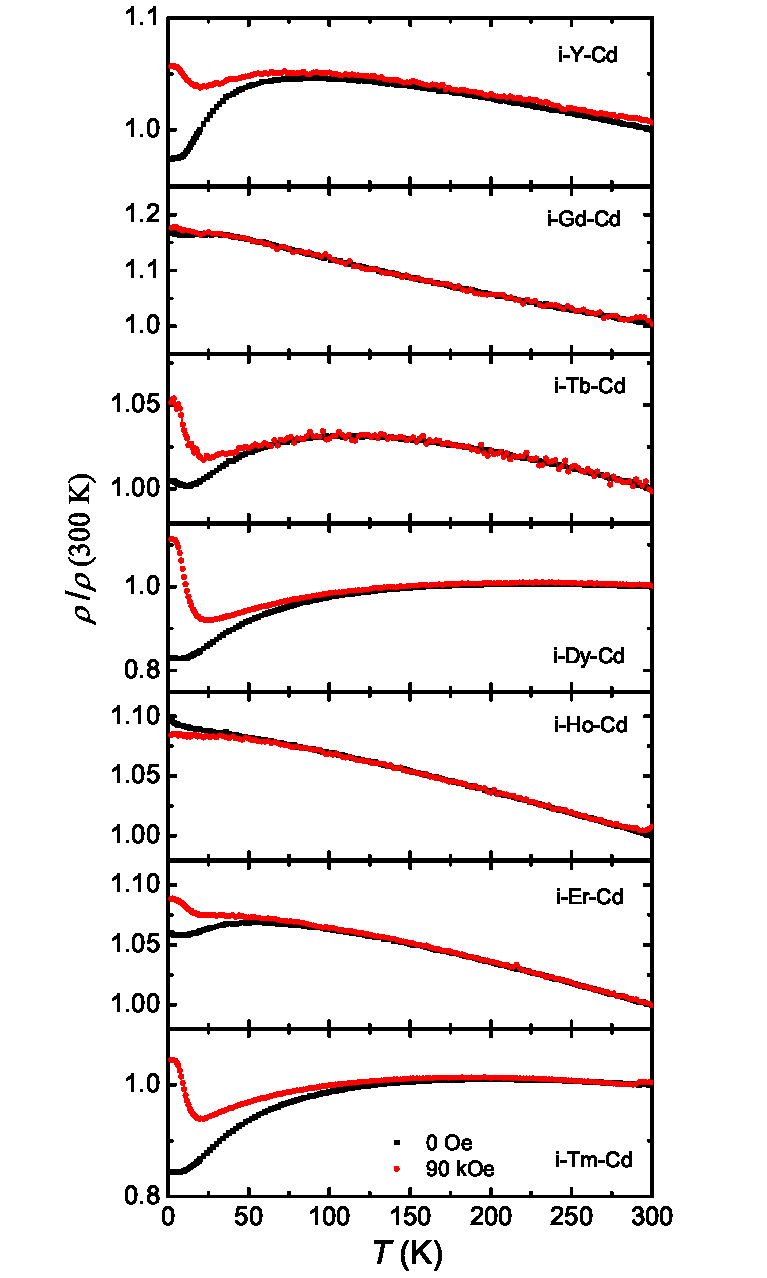}
\caption{(Color online) Normalized temperature-dependent resistance for i-$R$-Cd ($R$ = Y, Gd-Tm) at zero field (black) and at 90 kOe (red). The value of resistivity is about 300 $\mu\Omega$ cm at room-temperature.}
\label{resistivity}
\end{figure}

In Fig.~\ref{resistivity}, the normalized resistance data are measured in zero-field and 90 kOe for all i-$R$-Cd samples shown. The room-temperature resistivity of i-$R$-Cd quasicrystals is approximately 300 $\mu\Omega$ cm. Compared with other quasicrystal systems, this is close to that in i-$R$-Mg-Zn\cite{Fisher99} and i-Yb-Cd\cite{Pope01} and is an order of magnitude smaller than in Al-Pd-Mn and Al-Cu-Fe\cite{Janot1997, Swenson02}. If an assumption of a low intrinsic magnetoresistance of quasicrystals is made, it seems that i-Gd-Cd and i-Ho-Cd might be the best representation for single phase quasicrystal behavior. The resistance if i-Gd-Cd and i-Ho-Cd tends to increase with decreasing temperature. At 1.8 K, the magnetoresistance is small, and negative for i-Ho-Cd and positive for i-Gd-Cd. In other quasicrystal systems, both positive and negative magnetoresistance has been observed\cite{Akiyama01}. Besides the negative slope of zero-field resistance at room temperature, the resistance of i-$R$-Cd also exhibits a broad dome at lower temperature, for example, $\sim$100 K for i-Tb-Cd and $\sim$40 K for i-Gd-Cd. Although this feature might be affected by metallic Cd, similar dome-shape resistances had been observed in many other quasicrystals as well and was explained by weak localization with strong spin-orbit coupling\cite{Janot1997, Pope01, Swenson02, Akiyama01, Fukuyama81} and competing inelastic scattering in the presence of weak localization\cite{Jaiswal06}. It worth pointing out that the existence of such a broad maxima was shown to be closely related to the sample preparation method and the sample quality\cite{Fisher99,Kashimoto98}. In addition, at low temperatures, a small upturn or saturation in resistance exists, which sometimes appears as well in other aforementioned quasicrystals that exhibit a broad dome in resistance. The temperature at which the upturn or saturation occurs does not match dc magnetization features.

It is important to point out, though, that no sharp feature can be found in any of the data sets shown in  Fig.~\ref{resistivity} that can be associated with long-range magnetic ordering. This is consistent with the lack of $\lambda$-shaped features in specific heat measurements. For spin-glass systems, this is usually the case.

\bibliographystyle{apsrev4-1}

\end{document}